# How reliable are unsupervised author disambiguation algorithms in assessment of research organization performance?

Giovanni Abramo,[1] Ciriaco Andrea D'Angelo[2]

[1] *giovanni.abramo@iasi.cnr.it*
Laboratory for Studies in Research Evaluation, Institute for System Analysis and Computer Science (IASI-CNR), National Research Council of Italy (Italy)
ORCID: 0000-0003-0731-3635

[2] *dangelo@dii.uniroma2.it*
Department of Engineering and Management, University of Rome "Tor Vergata" (Italy)
&
Laboratory for Studies in Research Evaluation, Institute for System Analysis and Computer Science (IASI-CNR), National Research Council of Italy (Italy)
ORCID: 0000-0002-6977-6611

**Abstract**
The paper examines extent of bias in the performance rankings of research organisations when the assessments are based on unsupervised author-name disambiguation algorithms. It compares the outcomes of a research performance evaluation exercise of Italian universities using the unsupervised approach by Caron and van Eck (2014) for derivation of the universities' research staff, with those of a benchmark using the supervised algorithm of D'Angelo, Giuffrida, and Abramo (2011), which avails of input data. The methodology developed could be replicated for comparative analyses in other frameworks of national or international interest, meaning that practitioners would have a precise measure of the extent of distortions inherent in any evaluation exercises using unsupervised algorithms. This could in turn be useful in informing policy-makers' decisions on whether to invest in building national research staff databases, instead of settling for the unsupervised approaches with their measurement biases.

**Keywords**
*Research assessment; evaluative scientometrics; author name disambiguation; FSS; ORP; CWTS; universities; Italy.*

**Acknowledgement**
We are indebted to the Centre for Science and Technology Studies (CWTS) at Leiden University for providing us with access to the in-house WoS database from which we extracted data at the basis of our elaborations.

# 1. Introduction

The tools of performance assessment play a fundamental role in the strategic planning and analysis of national and regional research systems, member organizations and individuals. At the level of research organizations, assessment serves in identifying fields of strength and weakness, which in turn inform competitive strategies, organizational restructuring, resource allocation, and individual incentive systems. For regions and countries, knowledge of strengths and weaknesses relative to others, and also the comparative performances of one's own research institutions, enables formulation of informed research policies and selective allocation of public funding across fields and institutions. By assessing performance before versus after, institutions and governments can evaluate the effectiveness of their strategic actions and implementation of policy. The communication of the results from research assessment exercises, applied at any level, stimulates the assessed subjects towards continuous improvement. Such assessments also serve in reducing information asymmetries between the suppliers (researchers, institutions, territories) and the end users of research (companies, students, investors). At the macro-economic level, this yields twofold beneficial results, resulting in a virtuous circle: i) in selecting research suppliers, users can make more effective choices; and ii) suppliers, aiming to attract more users, will be stimulated to improve their research production. The reduction of asymmetric information is also beneficial within the scientific communities themselves, particularly in the face of the increasing challenges of complex interdisciplinary research, by lowering obstacles among prospective partners as they seek to identify others suited for inclusion in team-building.

Over recent years, the stakeholders of research systems have demanded more timely assessment, capable of informing in ever more precise, reliable and robust manner. Bibliometrics, and in particular evaluative bibliometrics, has the great advantage of enabling large-scale research evaluations with levels of accuracy, costs and timescales far more advantageous than traditional peer-review (Abramo, D'Angelo, & Reale, 2019), as well as possibilities for informing small-scale peer-review evaluations. For years, in view of the needs expressed by policy makers, research managers and stakeholders in general, scholars have continuously improved the indicators and methods of evaluative bibliometrics. In our opinion, however, the factor holding us back from a great leap forward is the lack of input data, which in almost all nations has been very difficult to assemble.

In all production systems, the comparative performance of any unit is always given by the ratio of outputs to inputs. In the case of research systems, the inputs or production factors consist basically of labor (the researchers) and capital (all resources other than labor, e.g. equipment, facilities, databases, etc.). For any research unit, therefore, comparison to another demands that we are informed of the component researchers, and the resources they draw on for conducting their research. In addition, bias in results would occur unless also informed of the prevailing research discipline of each researcher, since output is in part a function of discipline (Sorzano, Vargas, Caffarena-Fernández, & Iriarte, 2014; Piro, Aksnes & Rørstad, 2013; Lillquist & Green, 2010; Sandström & Sandström, 2009; Iglesias & Pecharromán, 2007; Zitt, Ramanana-Rahari, & Bassecoulard, 2005): scholars of blood diseases, for example, publish an average of about five times as much as scholars of legal medicine (D'Angelo & Abramo, 2015). Finally, the measure of the researcher's contribution to each scientific output should also take into account the number of co-authors, and in some cases their position in the author list (Waltman & Van

Eck, 2015; Abramo, D'Angelo, & Rosati, 2013; Aksnes, Schneider, & Gunnarsson, 2012; Huang, Lin, & Chen, 2011; Gauffriau & Larsen, 2005; van Hooydonk, 1997; Rinia, De Lange, & Moed, 1993).

Yet for many years, regardless of all these requirements, organizations have regularly published research institution performance rankings that are coauthor-, size- and field-dependent, among which the most renowned would be the *Academic Ranking of World Universities (ARWU)*,[1] issued by Shanghai Jiao Tong University, the *Times Higher Education World University Rankings*,[2] and the QS World University Rankings.[3] Despite the strong distortions in these rankings, many decision-makers persist in giving them serious credit. One of the most recent gestures of the sort came in May 2022, when the British government, intending well for those seeking immigration but without a job offer, offered early-career "High Potential Individuals" the possibility of a visa, subject to graduation with the past five years from an eligible university: meaning any university placing near top of the above – highly distorted – rankings.[4]

To get around the obstacle of missing input data, some bibliometricians have seen a solution in the so-called "size-independent" indicators of research performance - among these the mean normalized citation score (MNCS), proposed by the CWTS of the University of Leiden (Waltman et al., 2011; Moed, 2010). As we have pointed out, however, these indicators have strong limitations (Abramo & D'Angelo, 2016a, 2016b), and result in performance scores and ranks that are different from those obtained using other indicators, such as FSS (fractional scientific strength),[5] which do account for inputs, albeit with certain unavoidable assumptions. But the FSS indicator has thus far been applied in only two countries, both with advantages of government records on inputs: extensively, in Italy, for the evaluation of performance at the level of individuals (Abramo & D'Angelo, 2011) and then aggregated at the levels of research field and university (Abramo, D'Angelo, & Di Costa, 2011), and to a lesser extent in Norway, with additional assumptions (Abramo, Aksnes, & D'Angelo, 2020).

For policy-makers and administrators, but also all interested others, the question then becomes: "in demanding and/or using large-scale assessments of the positioning of research institution performance, what margin of error is acceptable in the measure of their scores and ranks?" To give an idea of the potential margins of error: a comparison of research-performance scores and ranks of Italian universities by MNCS and FSS revealed that 48.4% of universities shifted quartiles under these two indicators, and that 31.3% of universities in the top quartile by FSS fell into lower quartiles by MNCS (Abramo & D'Angelo, 2016c).

Italy is an almost completely unique case in the provision of the data on research staff at universities, as necessary for institutional performance evaluation. Here, at the close of each year, the Ministry of University and Research (MUR) updates a database of all university faculty members, listing the first and last names of each researcher, their gender, institutional affiliation, field classification and academic rank.[6] The Norwegian

---

[1] https://www.shanghairanking.com/rankings/arwu (last accessed 20/06/2022).
[2] https://www.timeshighereducation.com/world-university-rankings (last accessed 20/06/2022).
[3] *https://www.topuniversities.com/university-rankings/world-university-rankings/2022* (last accessed 20/06/2022).
[4] https://www.gov.uk/government/publications/high-potential-individual-visa-global-universities-list.
[5] A thorough explanation of the theory and assumptions underlying FSS can be found in Abramo and D'Angelo (2014), and in the more recent Abramo, Aksnes, and D'Angelo (2020).
[6] http://cercauniversita.cineca.it/php5/docenti/cerca.php, last accessed on 20/06/2022.

Research Personnel Register also offers a useful database of statistics,[7] including notation of the capital cost of research per man-year aggregated at area level, based on regular reports from the institutions to the Nordic Institute for Studies in Innovation, Research and Education (NIFU).[8]

The challenge facing practitioners is then how to apply output-input indicators of research performance aligned with microeconomic theory of production (like FSS), in all those countries where databases of personnel are not maintained. One possibility is to trace the research personnel of the institutions indirectly, through their publications, using bibliographic repertories such as Scopus or Web of Science (WoS), and referring exclusively to bibliometric metadata, apply algorithms for disambiguation of authors' names and reconciling of the institutions' names.

Computer scientists and bibliometricians have developed several unsupervised algorithms for disambiguation, at national and international levels (Rose & Kitchin, 2019; Backes, 2018; Hussain & Asghar, 2018; Zhu et al., 2017; Liu et al., 2015; Caron & van Eck, 2014; Schulz et al., 2014; Wu, Li, Pei, & He, 2014; Wu & Ding, 2013; Cota, Gonçalves, and Laender, 2007). The term "unsupervised" signifies that the algorithms operate without manually labelled data, instead approaching the author-name disambiguation problem as a clustering task, where each cluster would contain all the publications written by a specific author. Tekles & Bornmann (2020), using a large validation set containing more than one million author mentions, each annotated with a Researcher ID (an identifier maintained by the researchers), compared a set of such unsupervised disambiguation approaches. The best performing algorithm resulted as the one by Caron and van Eck (2014), hereinafter "CvE". As discussed above, however, the conduct of performance comparisons at organizational level requires more than just precision in unambiguously attributing publications to each author. At that point we also need precise identification of the research staff of each organization,[9] the fields of research, etc. And so if the aim is to apply bibliometrics for the comparative evaluation of organizational research performance, the goodness of the algorithms should be assessed on the basis of the precision with which they actually enable measurement of such performance.

To answer the research question, we therefore compare measures of the research performance of universities in the Italian academic system, which arise from the application of the previously conformed CvE unsupervised algorithm, with those arising from the use of the supervised algorithm by D'Angelo, Giuffrida, & Abramo (2011), hereinafter "DGA". Over more than a decade, this algorithm has been applied by the authors for feeding and continuous updating of the Public Research Observatory (ORP) of Italy, a database derived under license from Clarivate Analytics' WoS Core Collection. It indexes the scientific production of Italian academics at individual level, achieving 97% harmonic average of precision and recall (F-measure),[10] thanks to the operation of the DGA algorithm, which avails of a series of "certain" metadata available in the MUR database on university personnel, including their institutional affiliation, academic rank,

---

[7] https://www.nifu.no/en/statistics-indicators/4897-2/, last accessed on 20/06/2022
[8] http://www.foustatistikkbanken.no/nifu/?language=en, last accessed on 20/06/2022.
[9] The affiliation in the byline, in some cases multiple, is not always reliable to unequivocally identify the organisation to which the author belongs.
[10] The most frequently used indicators to measure the reliability of bibliometric datasets are precision and recall, which originate from the field of information retrieval (Hjørland, 2010). Precision is the fraction of retrieved instances that are relevant while recall is the fraction of relevant instances that are retrieved.

years of tenure, field of research, and gender (for details see D'Angelo, Giuffrida, & Abramo, 2011),

Given the maturity of the ORP, developed and refined year by year through the manual correction of the rare false cases, it can be considered a reliable benchmark against which to measure the deviations referable to an evaluation conducted using CvE. The deviations, as we shall see in some detail, are attributable to causes further than simply its lesser abilities in correctly disambiguating authorship. The aim of our work, however, is not to criticize CvE, but to give bibliometricians, practitioners, and especially decision makers, an idea of the extent of distortions in the research performance ranks of research institutions at overall and area level when forced to use unsupervised algorithms of this kind, rather than supervised ones based on research staff databases, such as DGA.

The paper is organized as follows: Section 2 presents the methodology and describes the data and methods used. In Section 3 we show the results of the analysis. Section 4 concludes, summarizing and also commenting the results, particularly for practitioners and scholars who may wish to replicate the exercise in other geographical and institutional frameworks.

## 2. Data and methods

The assessment of the comparative research performance of an organization cannot proceed without survey of the scientific activity of its individual researchers, since evaluations that operate directly at an aggregate level, without accounting for the sectoral distribution of input, produce results with unacceptable error (Abramo & D'Angelo, 2011). Analyses at micro level, however, presuppose precise knowledge of the research staff of the organization, as well as for all "competitor" organizations eligible for comparative evaluation. As explained above, the current study aims to compare the outcomes of the evaluation of research performance by Italian universities, based on two different bibliometric datasets:

- The first one, hereinafter "ORP", relying on the DGA supervised heuristic approach, which "integrates" the Italian National Citation Report (indexing all WoS articles by those authors who indicated Italy as affiliation country), with data retrieved from the database maintained by the MUR,[11] indexing the full name, academic rank, research field and institutional affiliation of all researchers at Italian universities, at the close of each year.
- The second one, hereinafter "CWTS", relying on the CvE unsupervised approach, a rule-based scoring and oeuvre identification method for disambiguation of authors used for the WoS in-house database of the Centre for Science and Technology Studies (CWTS) at Leiden University.

Much fuller descriptions of the DGA and CvE approaches can be found in D'Angelo and van Eck (2020).

In ORP:
- a priori, the availability of MUR data allows precise knowledge of the members of research staff of national universities;
- the census of their scientific production is then carried out by applying the DGA algorithm to the Italian WoS publications.

---

[11] http://cercauniversita.cineca.it/php5/docenti/cerca.php, last accessed 20/06/2022.

In CWTS the two properties are inseparable, and the research staff of an organization is derived directly in extracting bibliometric data from WoS: the CvE algorithm associates clusters of publications with a proto-individual on the basis of the similarity of the publications' metadata. In Table 1 we give as an example the information that the algorithm associates with the cluster referred to the second author of the present work (Ciriaco Andrea D'Angelo).

*Table 1: Description of the output of the CvE approach for one of the authors of this paper*

| Field | Value |
| --- | --- |
| cluster_id | 56122902 |
| n_pubs | 129 |
| first_year | 1996 |
| last_year | 2020 |
| "Academic" age | 24 |
| full_name | d'angelo, ca |
| last_name | d'angelo |
| first_name | ciriaco andrea |
| email | dangelo@dii.uniroma2.it |
| organization | univ roma tor vergata |
| city | rome |
| country | italy |
| orcid | 0000-0002-6977-6611 |
| researcherid | J-8162-2012 |

In particular, each proto-individual, uniquely identified by means of a "cluster_id" (56122902 in the example in Table 1), is associated with an "organisation" (univ roma tor vergata) on the basis of the most recurrent affiliation in the publications assigned to them, as well as an email (dangelo@dii.uniroma2.it) on the basis of the same criterion.

The identification of the research staff of a university requires analysis of all possible variant terms representing it as the prevailing "organisation" of the proto-individuals indexed in CWTS. The Italian case also allows the alternative of referring to "email", since the national academic system provides for a standardised web domain (e.g. "uniroma2.it" for Roma "Tor Vergata", "unimi.it" for University of Milano, "unipa.it" for University of Palermo). In this regard, the box below shows the query related to the extraction of clusters for University of Rome 'Tor Vergata'.[12]

```
([organization] = ("state univ rome tor vergata" OR "tor vergata univ" OR "tor vergata univ rome" OR "univ roma tor vergata" OR "univ roma tor vergata 2" OR "univ tor vergata")

OR [email] like '%uniroma2.it')

AND [country]='italy'
```

For reasons of robustness, after the first data extraction of the current analyses, we eliminated universities for which the procedure had identified fewer than 30 clusters (typically telematic or predominantly humanistic universities). For the remaining universities (65 in all), we performed subsequent quality control and tuning operations. The combination of the two conditions ([organisation] OR [email]) in fact makes it possible to maximise the recall of the procedure, but also inevitably generates false positives, i.e. retrieval of subjects that do not actually belong to the institution in question. Such "incoherent" clusters, include those:

- where the "organisation" remains ambiguous, or does not refer to a recognised university;
- with email not referring to a university;
- with email and organisation referring to different universities.

---

[12] It can be a formidable task, looking at all bibliometric addresses, to identify the variants of "organization" attributable to a single institution. In the present case, we are dealing with 90 universities and manually scanning the 4787 total name variants "associated" with their official emails. Practitioners facing larger numbers or constraints on resources could decide to manually check only the first "n" in terms of frequency. The six "organization" variants in the case shown in the box, for example, account for 95% of all clusters linked with a "%uniroma2.it" email.

These control operations, for example, lead to the exclusion of the first author of the current article (Giovanni Abramo) from the dataset, for whom the initial extraction gives an assignment to the University of Rome 'Tor Vergata' because of the email (giovanni.abramo@uniroma2.it), even though his organization is "natl res council italy", i.e. a non-academic research body.

Furthermore, to exclude "occasional" and terminated researchers, we impose:

- an "academic age" of at least 4 years (given by the difference between the years of the most recent and the first publications);
- the most recent publication of the cluster no earlier than 2020.

Finally, conflicts involving distinct clusters, but sharing the same "orcid" organization ID or email, are resolved manually.

To assess the yearly average performance of each researcher in a period of time, we recur to the Fractional Scientific Strength ($FSS_R$) indicator of research productivity, defined as:

$$FSS_R = \frac{1}{t}\sum_{i=1}^{N}\frac{c_i}{\bar{c}}f_i$$

[1]

where:
t = number of years of work of the researcher in the period under observation;[13]
$N$ = number of publications[14] of the researcher in the period under observation;
$c_i$ = citations received by publication *i*;
$\bar{c}$ = average of distribution of citations received by all publications in the same year and WoS subject category (SC) of publication *i*;
$f_i$ = fractional contribution of the researcher to publication *i*, given by the inverse of the number of co-authors in the byline.

The indicator is calculated over the period 2015-2019, with the citation count at week 13 in 2021. Performance measured at the individual level is then aggregated to obtain the performance of a university ($FSS_U$) at the SC, area[15] or overall level. In formulae:

---

[13] For researchers in CWTS database we assume t=5 for all.
[14] We consider all publications indexed in the WoS *core collection* (excluding ESCI) with document types: articles, reviews, letters, proceedings.
[15] SCs are classified and grouped into areas according to a system previously published on the webpage of the ISI Journal Citation Reports. This page is no longer available at the current Clarivate site. It should be noted that all SCs are assigned to only one area.

$$FSS_U = \frac{1}{RS_U} \sum_{j=1}^{RS_U} \frac{FSS_{R_j}}{\overline{FSS_R}}$$

[2]

where:

$RS_U$ = research staff of the university unit, in the observed period;
$FSS_{R_j}$ = productivity of researcher *j* in the unit;
$\overline{FSS_R}$ = national average productivity of all productive researchers in the same SC of researcher *j*.

Prior to any aggregation of data by university unit, it is absolutely necessary that individual performance be scaled against the expected value of the reference SC, but this requires an "SC classification" of each researcher. For this purpose, we used the WoS classification scheme, assigning the prevailing SC as follows:
- for the CWTS evaluation, with reference to the researcher's entire scientific production in WoS; in uncertain cases (researcher with multiple prevailing SCs) randomly among those with a higher frequency;
- for the ORP evaluation, with reference to the researcher's scientific production 2001-2019; in uncertain cases (researcher without publications or with multiple prevalent SCs) the one with the highest incidence relative to the SDS[16]-SC pair.

In review, we note that the performance indicator is calculated in the same way for both the ORP and CWTS datasets.[17] Due to its derivation, however, in comparison to the ORP, the CWTS dataset has two limitations. First, it does not contain those resear*che*rs who have never published in the period under observation, as they cannot be identified from a bibliographic repertory, although they must contribute to the evaluation as they represent a research cost. Second, in the period under observation, the years on staff for each researcher remain unknown, so that if one presents publications in CWTS only over 2017-2019, for example, it is unknown whether the lack of production in years 2015-2016 was because they were not on staff. Instead, the information is known in ORP.

For reasons of significance, the analysis excludes researchers in SCs belonging to "Art and Humanities" and "Law, political and social sciences", where the coverage of bibliographic repertories is scarce (Hicks, 1999; Larivière, Archambault, Gingras, Vignola-Gagné, 2006; Aksnes & Sivertsen, 2019). The analysis is then restricted to the researchers of SCs pertaining to the STEM and Economics, but also excluding subjects classified in "Multidisciplinary Sciences", and again for the sake of significance, SCs with fewer than 10 observations in both datasets. Table 2 shows the resulting breakdown by area for the two datasets.

---

[16] In the Italian university system all professors are classified in one and only one field (named the scientific disciplinary sector, SDS, 370 in all).
[17] This does not mean that the value is necessarily identical for the subjects in both datasets, since: i) the CvE and DGA algorithms are not free from error in attributing to a given author the publications he or she has actually produced; and ii) in ORP, the value of "t" may differ from 5.

*Table 2: Number of researchers in the two datasets for analysis, by area\**

| Area | No. of obs dataset ORP | No. of obs dataset CWTS | Delta |
|---|---|---|---|
| Biology | 4928 | 7987 | +62.1% |
| Biomedical research | 2891 | 7217 | +149.6% |
| Chemistry | 1606 | 2735 | +70.3% |
| Clinical medicine | 6205 | 13444 | +116.7% |
| Earth and space sciences | 1937 | 3039 | +56.9% |
| Economics | 3784 | 1540 | -59.3% |
| Engineering | 5554 | 7873 | +41.8% |
| Mathematics | 2055 | 1805 | -12.2% |
| Physics | 2617 | 3793 | +44.9% |
| Psychology | 412 | 475 | +15.3% |
| Total | 31989 | 49908 | +56.0% |

\* *The counts exclude SCs with less than 10 observations in both datasets*

Overall, the CWTS evaluation is based on 49,908 subjects, 56% more than the 31,989 in ORP. Since the ORP dataset is the benchmark, we can reasonably consider that, in CWTS, the number of false positives (researchers assigned to a university although not officially part of the research staff) is significantly higher than the number of false negatives (researchers not assigned to a university although part of the research staff).

The data in Table 2 indicate, however, that the over-representation of CWTS compared to ORP is not identical between the areas. Biomedical research and Clinical medicine have the largest deviations (+149.6% and +116.7% respectively). In Economics, on the contrary, we have 2.5 times more observations in ORP than in CWTS (3784 vs 1540).

## 3. Results

### 3.1 The distributions of performance at individual level

First, we analyse the differences between the two datasets in the distributions of performance at individual level ($FSS_R$). We report the results of three SCs, exemplary of different types of cases. Table 3 shows the descriptive performance statistics for the authors of "Engineering, manufacturing"; "Dermatology"; "Statistics & probability"; Figure 1 shows the box plots of the distributions for the last two SCs.

For "Engineering, manufacturing", the number of observations in the two datasets results as virtually identical (145 vs 143), and the distributions seem almost superimposable, with practically identical mean/median and dispersion values. Moreover, the maximum values coincide: referring to the same subject, Professor Francesco Lambiase of the University of L'Aquila. Overall, this accordance occurs in no less than 39 SCs out of the total 160. It can also be noted that in ORP, the minimum value of the distribution is equal to 0, indicating that in the evaluation of the SC using this dataset, we find at least one unproductive researcher (FSS=0), which does not happen with CWTS (where the minimum recorded is equal to 0.018). Overall, this circumstance is detectable in only two SCs besides "Engineering, manufacturing". In contrast, there are 31 SCs where CWTS registers at least one unproductive (FSS=0), versus ORP finding none.

In "Dermatology", the numerosity of researcher observations the two datasets is very different, with 258 in CWTS compared to 108 in ORP. In this case the distribution of CWTS performance seems decidedly more shifted to the left than its counterpart in ORP, with lower values both in terms of mean (0.453 vs 0.942) and median (0.139 vs 0.601). In terms of percentiles, the CWTS distribution also shows systematically lower values than the ORP distribution, with the sole exception of the maximum (6.639 vs 6.488).

The situation is diametrically opposed in "Statistics & probability", an SC where the CWTS observations (277) are well below the ORP observations (442). In this case, the CWTS performance distribution appears decisively shifted to the right compared to that for ORP, with higher values of mean (0.353 vs 0.306), median (0.212 vs 0.151), and all percentiles.

*Table 3: Descriptive statistics of the distribution of $FSS_R$ for researchers of three SCs, comparing ORP and CWTS datasets*

|  |  | Engineering, manufacturing | | Dermatology | | Statistics & probability | |
|---|---|---|---|---|---|---|---|
|  |  | ORP | CWTS | ORP | CWTS | ORP | CWTS |
|  | Obs | 145 | 143 | 108 | 258 | 442 | 277 |
|  | Mean | 0.944 | 1.034 | 0.942 | 0.453 | 0.306 | 0.353 |
|  | Std Dev. | 1.102 | 1.119 | 1.161 | 0.818 | 0.494 | 0.455 |
|  | Variance | 1.214 | 1.253 | 1.348 | 0.669 | 0.244 | 0.207 |
|  | Skewness | 3.267 | 2.958 | 2.190 | 3.471 | 4.887 | 5.248 |
|  | Kurtosis | 19.350 | 16.873 | 8.791 | 18.923 | 36.813 | 50.275 |
| Percentile | 1% | 0 | 0.018 | 0 | 0 | 0 | 0 |
|  | 5% | 0.023 | 0.056 | 0.021 | 0 | 0 | 0.006 |
|  | 10% | 0.056 | 0.134 | 0.053 | 0.004 | 0 | 0.028 |
|  | 25% | 0.234 | 0.357 | 0.151 | 0.030 | 0.049 | 0.095 |
|  | 50% | 0.683 | 0.656 | 0.601 | 0.139 | 0.151 | 0.212 |
|  | 75% | 1.317 | 1.357 | 1.203 | 0.443 | 0.387 | 0.454 |
|  | 90% | 2.012 | 2.342 | 2.670 | 1.372 | 0.749 | 0.856 |
|  | 95% | 2.733 | 3.227 | 3.041 | 2.162 | 0.996 | 1.048 |
|  | 99% | 5.164 | 4.422 | 5.499 | 3.455 | 2.403 | 1.666 |
|  | Max | 8.572 | 8.572 | 6.488 | 6.639 | 5.034 | 5.216 |

*Figure 1: Boxplots of $FSS_R$ distribution for authors in Dermatology (GA) and Statistics & probability (XY), comparing ORP and CWTS datasets*

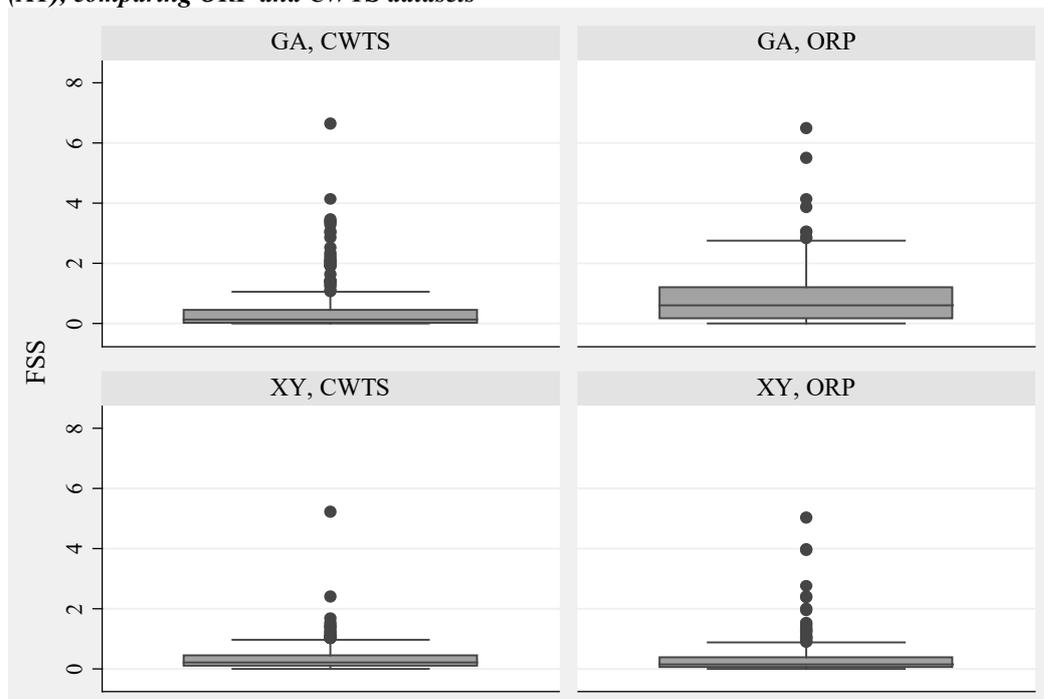

Figures 2 and 3 show the comparison of the mean and median values of the distributions for all 160 SCs, in the two datasets. At a glance, one can observe a greater number of cases in which the "central" values (mean and median) calculated in CWTS are lower than their counterparts recorded in ORP. Indeed, in 107 SCs out of a total of 160 (i.e. 67%), the mean value of the $FSS_R$ is higher in the ORP relative to the CWTS dataset, while for the medians, this occurs in 114 SCs out of 160 (71%).

Thus, in general, it would appear that the distribution of performance recorded in the ORP dataset is more rightward shifted than in the CWTS dataset. We wonder if this would be explained by the different composition of the two datasets and, in particular, the over-representation of CWTS compared to ORP. In fact, the Pearson $\rho$ correlation between the deviations in terms of the number of observations and the deviations between the mean $FSS_R$ values for the 160 SCs is -0.515 (-0.454 when considering the medians). Basically, in the SCs where the over-representation of the CWTS dataset compared to ORP is greater, the mean performance values in CWTS are significantly lower than those found in ORP, and vice versa. We can hypothesise that the so-called "false positives" have lower average performance values than the "true positives". Or, if you like, "non-faculty" personnel, but with a bibliometrically prevalent university affiliation, have a lower average $FSS_R$ than the research staff truly on faculty at the universities in the same field of observation. This has an important implication concerning the use of CWTS for comparative performance assessment, in that it would evidently "penalise" those organisations (and areas within organizations) with a higher concentration of non-faculty personnel in their research staff.

*Figure 2: Distribution of the mean values of $FSS_R$ detected in the two datasets, for the 160 subject categories considered*

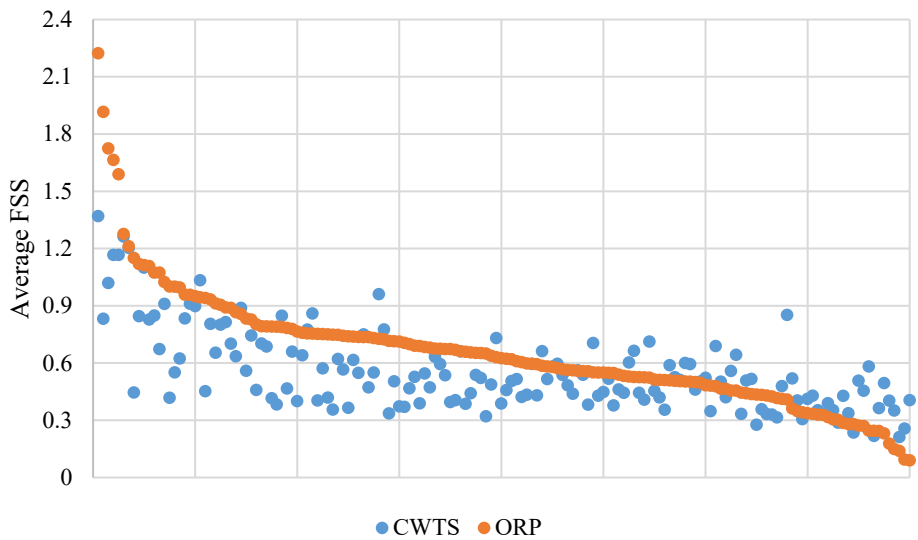

*Figure 3: Distribution of the median $FSS_R$ values detected in the two datasets, for the 160 subject categories considered*

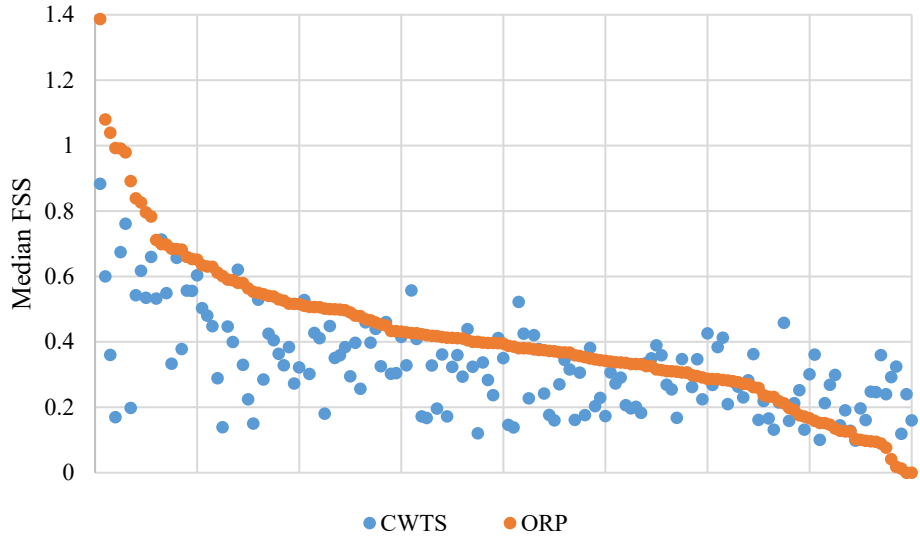

### 3.2 Evaluation of the universities' performance

We now move on to analyze the deviations between the two datasets in terms of the score and rank of the performance of the universities, through the $FSS_U$ indicator in formula [2]. Figure 4 shows the values measured at the overall level for the two datasets of the 65 universities with at least 30 observations of researchers. The dispersion of the values for the ORP dataset is greater than that for the CWTS (standard deviations 0.316 vs 0.175). Figure 5 instead shows the scatterplot of the values of the two indicators and evidences a strong correlation. The detailed values are shown in Table 4: the Università Vita - Salute San Raffaele and the Scuola Superiore S.Anna are at the top in both assessments. The top 11 universities between the two rankings vary by at most six positions (the case for the University of Padua and Politecnico di Milano). Thus, the evaluation by CWTS returns a top part of the ranking substantially similar to that resulting in ORP. Instead, the situation in the middle and lower part of the ranking is different, where LUISS stands out, which shifts 36 positions between the two rankings, and the University "Campus Bio-medico", which shifts 29. On the other hand, there are also five universities (University of Naples 'Parthenope'; University of Enna; University of the Mediterranean Studies of Reggio Calabria; University of Sannio; University of Teramo) which in the CWTS ranking gain between 31 and 34 positions compared to ORP. It is noticeable the gain in performance score by the bottom-ranked universities in the CWTS ranking. A possible interpretation is that the ORP performance of true positives is so low that false positives cannot help contributing to increase overall CWTS performance.

The magnitude and direction of these variations could be related to the differing numbers of researchers evaluated in the two modes. In fact, the percentage deviations of $FSS_U$ between the two datasets are significantly and negatively correlated with the percentage deviations of numerosity (Pearson $\rho = -0.526$). The same is true for the ranking jumps and the percentage deviations in numerosity between the two datasets (Pearson $\rho = -0.361$).

To summarise, compared to the ORP benchmark, the value of the performance recorded in CWTS (both absolute and relative) decreases as the concentration of "non-faculty" personnel in the research staff increases. Thus, the hypothesis is confirmed that conducted via CWTS, universities (and areas within universities) with a higher proportion of non-faculty research staff are actually penalised. This does not prevent the Vita - Salute San Raffaele University from coming out on top in both rankings, despite the fact that in the CWTS dataset there are 364 researchers associated with it compared to the 103 actual researchers recorded in ORP.

*Figure 4: Distribution of $FSS_U$ for universities in the CWTS and ORP datasets, at the overall level*

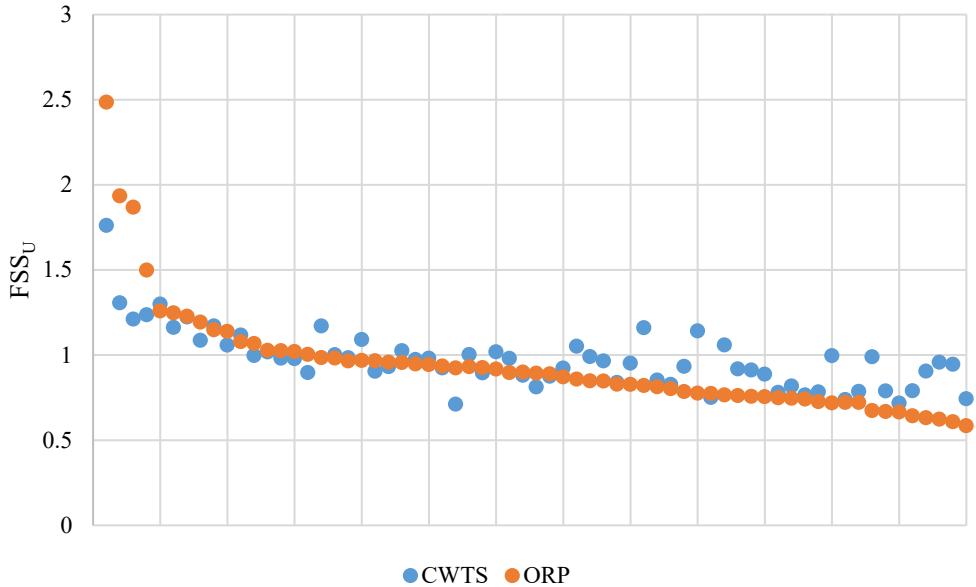

*Figure 5: Scatterplot of $FSS_U$ values for universities in the CWTS and ORP datasets, at the overall level*

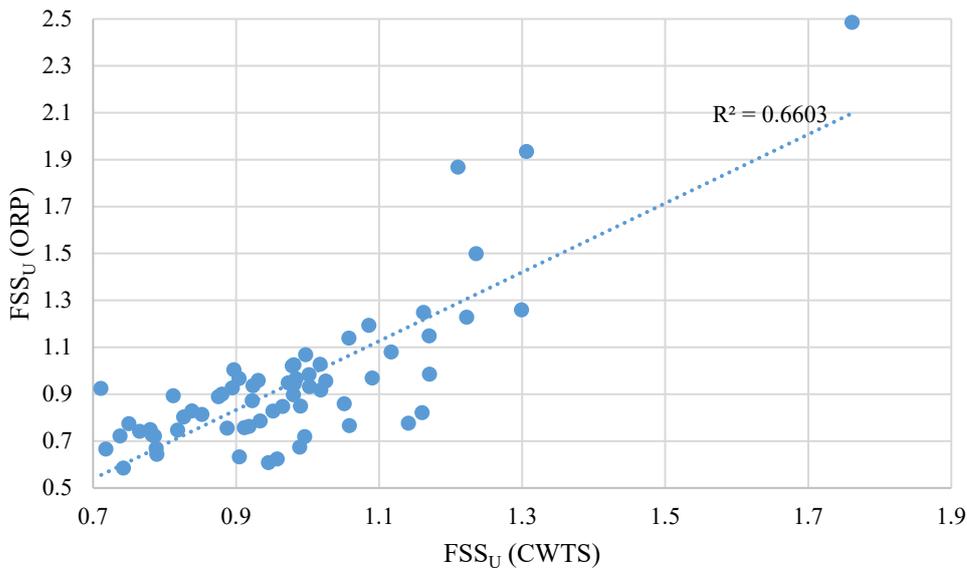

*Table 4: Performance score and rank of Italian universities, comparing CWTS and ORP datasets*

| Ateneo | CWTS | | | | ORP | | | | Δ Rank |
|---|---|---|---|---|---|---|---|---|---|
| | Obs | $FSS_U$ | Rank | Perc. | Obs | $FSS_U$ | Rank | Perc. | |
| Vita - Salute San Raffaele | 364 | 1.762 | 1 | 100 | 103 | 2.485 | 1 | 100 | 0 |
| Scuola Superiore S.Anna | 172 | 1.306 | 2 | 98 | 82 | 1.935 | 2 | 98 | 0 |
| SISSA | 128 | 1.210 | 6 | 92 | 67 | 1.868 | 3 | 97 | -3 |
| Libera Università di Bolzano | 103 | 1.236 | 4 | 95 | 91 | 1.499 | 4 | 95 | 0 |
| Commerciale Luigi Bocconi | 112 | 1.299 | 3 | 97 | 191 | 1.259 | 5 | 94 | +2 |
| Politecnico di Bari | 264 | 1.163 | 9 | 88 | 212 | 1.248 | 6 | 92 | -3 |
| Trento | 487 | 1.223 | 5 | 94 | 332 | 1.228 | 7 | 91 | +2 |
| Padova | 2845 | 1.086 | 14 | 80 | 1394 | 1.193 | 8 | 89 | -6 |
| Università di Salerno | 696 | 1.170 | 8 | 89 | 523 | 1.148 | 9 | 88 | +1 |
| Politecnico di Milano | 1339 | 1.058 | 16 | 77 | 993 | 1.139 | 10 | 86 | -6 |
| Napoli "Federico II" | 2405 | 1.117 | 12 | 83 | 1659 | 1.079 | 11 | 84 | -1 |
| Milano | 2724 | 0.998 | 23 | 66 | 1317 | 1.068 | 12 | 83 | -11 |
| Università di Pisa | 1622 | 1.018 | 20 | 70 | 908 | 1.027 | 13 | 81 | -7 |
| Verona | 838 | 0.981 | 29 | 56 | 421 | 1.025 | 14 | 80 | -15 |
| Firenze | 1959 | 0.979 | 31 | 53 | 983 | 1.020 | 15 | 78 | -16 |
| "Campus Bio-medico" | 257 | 0.897 | 45 | 31 | 107 | 1.004 | 16 | 77 | -29 |
| Istituto Univ. di Scienze Motorie | 50 | 1.171 | 7 | 91 | 42 | 0.985 | 17 | 75 | +10 |
| Perugia | 968 | 1.002 | 22 | 67 | 654 | 0.983 | 18 | 73 | -4 |
| Università di Catania | 976 | 1.091 | 13 | 81 | 708 | 0.968 | 19 | 72 | +6 |
| Torino | 2166 | 0.904 | 44 | 33 | 1121 | 0.966 | 20 | 70 | -24 |
| Politecnico di Torino | 988 | 0.984 | 27 | 59 | 658 | 0.966 | 21 | 69 | -6 |
| Università di Ferrara | 700 | 0.931 | 38 | 42 | 389 | 0.958 | 22 | 67 | -16 |
| Magna Grecia di Catanzaro | 290 | 1.026 | 18 | 73 | 162 | 0.956 | 23 | 66 | +5 |
| Pavia | 957 | 0.973 | 32 | 52 | 536 | 0.948 | 24 | 64 | -8 |
| Università di Bologna | 2889 | 0.982 | 28 | 58 | 1640 | 0.944 | 25 | 63 | -3 |
| Politecnica delle Marche | 707 | 0.924 | 39 | 41 | 436 | 0.936 | 26 | 61 | -13 |
| della Tuscia | 254 | 1.003 | 21 | 69 | 177 | 0.932 | 27 | 59 | +6 |
| Bergamo | 132 | 0.895 | 46 | 30 | 145 | 0.926 | 28 | 58 | -18 |
| LUISS | 32 | 0.711 | 65 | 0 | 51 | 0.924 | 29 | 56 | -36 |
| Università della Calabria | 627 | 1.019 | 19 | 72 | 473 | 0.918 | 30 | 55 | +11 |
| Cattolica del Sacro Cuore | 906 | 0.881 | 48 | 27 | 704 | 0.900 | 31 | 53 | -17 |
| Milano Bicocca | 1020 | 0.980 | 30 | 55 | 567 | 0.897 | 32 | 52 | +2 |
| Urbino "Carlo Bo" | 211 | 0.813 | 54 | 17 | 159 | 0.893 | 33 | 50 | -21 |
| del Salento | 382 | 0.876 | 49 | 25 | 298 | 0.889 | 34 | 48 | -15 |
| Università dell'Insubria | 360 | 0.923 | 40 | 39 | 241 | 0.872 | 35 | 47 | -5 |
| Messina | 810 | 1.052 | 17 | 75 | 642 | 0.859 | 36 | 45 | +19 |
| Roma Tre | 397 | 0.990 | 25 | 63 | 349 | 0.848 | 37 | 44 | +12 |
| Foggia | 275 | 0.965 | 33 | 50 | 211 | 0.847 | 38 | 42 | +5 |
| Genova | 1257 | 0.839 | 51 | 22 | 745 | 0.828 | 39 | 41 | -12 |
| dell'Aquila | 495 | 0.952 | 35 | 47 | 383 | 0.827 | 40 | 39 | +5 |
| Napoli "Parthenope" | 213 | 1.160 | 10 | 86 | 223 | 0.820 | 41 | 38 | +31 |
| Roma "La Sapienza" | 3414 | 0.852 | 50 | 23 | 2048 | 0.813 | 42 | 36 | -8 |
| Brescia | 702 | 0.827 | 52 | 20 | 410 | 0.802 | 43 | 34 | -9 |
| Roma "Tor Vergata" | 1158 | 0.934 | 37 | 44 | 849 | 0.785 | 44 | 33 | +7 |
| Enna | 38 | 1.141 | 11 | 84 | 56 | 0.776 | 45 | 31 | +34 |
| Cagliari | 798 | 0.750 | 61 | 6 | 560 | 0.774 | 46 | 30 | -15 |
| Mediterranea di Reggio Calabria | 181 | 1.059 | 15 | 78 | 159 | 0.765 | 47 | 28 | +32 |
| Gabriele D'Annunzio | 587 | 0.918 | 41 | 38 | 394 | 0.762 | 48 | 27 | +7 |
| Palermo | 1181 | 0.912 | 42 | 36 | 885 | 0.757 | 49 | 25 | +7 |
| Bari | 1193 | 0.888 | 47 | 28 | 852 | 0.755 | 50 | 23 | +3 |
| Parma | 895 | 0.780 | 59 | 9 | 610 | 0.749 | 51 | 22 | -8 |
| Trieste | 536 | 0.818 | 53 | 19 | 378 | 0.746 | 52 | 20 | -1 |
| Università di Camerino | 288 | 0.765 | 60 | 8 | 200 | 0.741 | 53 | 19 | -7 |
| Modena e Reggio Emilia | 888 | 0.783 | 58 | 11 | 537 | 0.726 | 54 | 17 | -4 |
| Siena | 752 | 0.786 | 57 | 13 | 392 | 0.722 | 55 | 16 | -2 |

|  | CWTS | | | | ORP | | | | |
|---|---|---|---|---|---|---|---|---|---|
| Ateneo | Obs | $FSS_U$ | Rank | Perc. | Obs | $FSS_U$ | Rank | Perc. | Δ Rank |
| Università Ca' Foscari Venezia | 198 | 0.738 | 63 | 3 | 184 | 0.721 | 56 | 14 | -7 |
| del Sannio | 154 | 0.996 | 24 | 64 | 142 | 0.719 | 57 | 13 | +33 |
| Teramo | 115 | 0.989 | 26 | 61 | 104 | 0.674 | 58 | 11 | +32 |
| Udine | 526 | 0.789 | 56 | 14 | 402 | 0.668 | 59 | 9 | +3 |
| Piemonte Orientale A. Avogadro | 178 | 0.718 | 64 | 2 | 234 | 0.666 | 60 | 8 | -4 |
| Sassari | 474 | 0.790 | 55 | 16 | 342 | 0.643 | 61 | 6 | +6 |
| del Molise | 166 | 0.905 | 43 | 34 | 133 | 0.632 | 62 | 5 | +19 |
| Seconda Napoli | 574 | 0.958 | 34 | 48 | 563 | 0.623 | 63 | 3 | +29 |
| Cassino | 146 | 0.946 | 36 | 45 | 152 | 0.608 | 64 | 2 | +28 |
| della Basilicata | 257 | 0.743 | 62 | 5 | 222 | 0.585 | 65 | 0 | +3 |

Depending on the particular application one has in mind (e.g., for a policy-maker, program administrator), the sensitivity of any measuring instrument will be more or less critical. In the case of the data in Table 4, a variation to the second decimal place in the values of $FSS_U$ results in some major jumps in ranking. This certainly prompts thinking that a less precise ranking would be reasonable, such as by performance classes, e.g. by quartiles, which are in fact used in some national research evaluation exercises. In Table 5 we report the ranking of the universities' performance in the two datasets, by quartiles. The data show that 36 out of 65 universities are ranked in the same quartile in the two modes (in the main diagonal). The remaining 29 show a jump of at least one quartile. Of these:
- 13 have a better ranking based on CWTS than ORP (above the main diagonal);
- 16 have a worse ranking based on CWTS than ORP (below the main diagonal).

The nine universities shown in Table 6 experience a two-quartile jump between the two rankings. No university experiences a three-quartile jump, i.e. top to bottom or vice versa.

*Table 5: Performance quartiles of the universities evaluated using CWTS and ORP datasets*

|  |  | ORP | | | |
|---|---|---|---|---|---|
|  |  | I | II | III | IV |
| CWTS | I | 12 | 1 | 4 | 0 |
|  | II | 4 | 8 | 2 | 2 |
|  | III | 1 | 5 | 6 | 4 |
|  | IV | 0 | 2 | 4 | 10 |

*Table 6: Universities showing two-quartile ranking jumps between the two datasets*

| Ateneo | Quartile CWTS | Quartile ORP |
|---|---|---|
| Messina | 1 | 3 |
| Napoli "Parthenope" | 1 | 3 |
| Enna | 1 | 3 |
| Mediterranea di Reggio Calabria | 1 | 3 |
| Sannio | 2 | 4 |
| Teramo | 2 | 4 |
| "Campus Bio-medico" | 3 | 1 |
| LUISS | 4 | 2 |
| Urbino "Carlo Bo" | 4 | 2 |

What was just seen in section 3.2 at the overall level is repeated at the area level. Figure 6 shows the scatterplot of $FSS_U$ calculated in the two modes, at this level of aggregation. Table 7 presents the correlation between score and ranking of the 65 universities evaluated by area. At the score level, the Pearson coefficient recorded in the two assessments at the overall level is 0.816. The figure for the individual areas is never less than 0.5 with the sole exception of Psychology, which is an area with a very low number of assessable universities (only 9). The correlation coefficients between the ranks are slightly lower, i.e. 0.686 at overall level. This confirms the exception of Psychology, an area in which the two assessment modes lead to results that are completely noncorrelated. Chemistry is confirmed as the area with the most closely aligned ranks between the two modes (Spearman 0.830) followed by Earth and Space Sciences (0.747) and Economics (0.712).

*Figure 6: Distributions of performance evaluated in the two modes, by area*

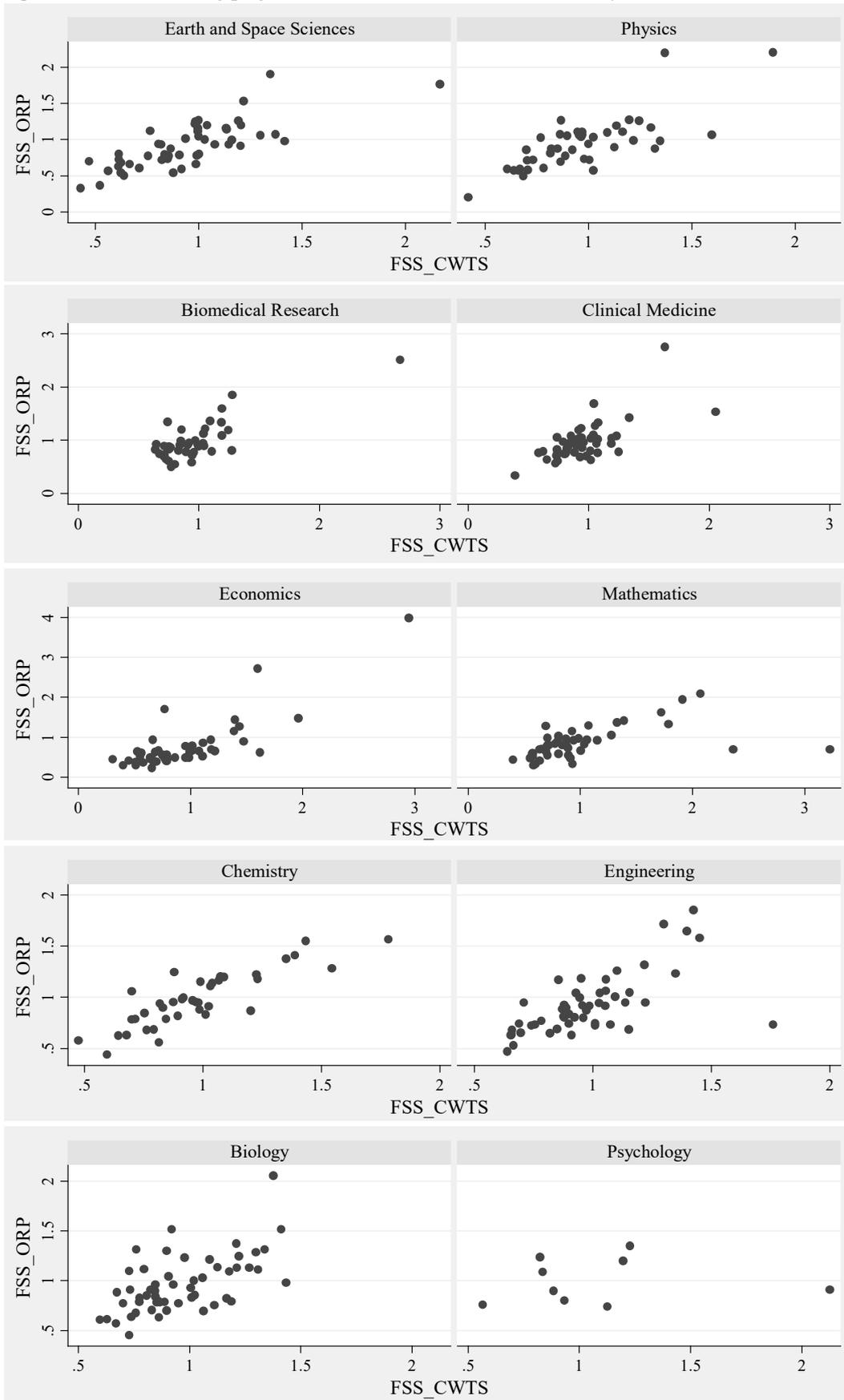

*Table 7: Correlation between score and ranking of the 65 Italian universities, evaluated in the two modes, by area*

| Discipline | No. of obs (evaluated universities) | Pearson ρ | Spearman ρ |
|---|---|---|---|
| Biology | 54 | 0.602 | 0.576 |
| Biomedical Research | 42 | 0.789 | 0.537 |
| Chemistry | 39 | 0.851 | 0.824 |
| Clinical Medicine | 47 | 0.681 | 0.580 |
| Earth and Space Sciences | 50 | 0.775 | 0.747 |
| Economics | 48 | 0.808 | 0.712 |
| Engineering | 52 | 0.657 | 0.673 |
| Mathematics | 48 | 0.504 | 0.607 |
| Physics | 44 | 0.760 | 0.702 |
| Psychology | 9 | 0.072 | 0.233 |
| Overall | 65 | 0.813 | 0.686 |

## 4. Discussion and conclusions

Evaluative bibliometricians have been engaged for years in the continuous improvement of indicators and methods for evaluating the scientific activities of individuals, organisations, and national and territorial research systems. The major obstacle hindering a leap ahead in evaluation techniques is the lack of input data. Yet a basic principle is that the proper evaluation of the performance of any subject at any level, including in research performance, requires data on inputs. In particular, assessing the performance of universities requires knowledge of the individual researchers working within them and the resources they use to carry out their research projects. For these reasons, the question of the reliability of author disambiguation algorithms in identifying the true publications of each observed subject must be combined with that of their ability in identifying the research staff of the home institutions, including their placement in disciplinary fields.

In this work we evaluated the goodness-of-fit of the CvE unsupervised author-name disambiguation algorithm in measuring the performance scores and ranks of Italian universities, operating through direct processing of bibliometric data for deduction of the research staff (and thus the input data) of each university. The validation was carried out through the comparison with the supervised DGA algorithm, based on the a priori knowledge of the research staff truly in post in each national university.

The results of the comparison showed that the unsupervised approach overestimates the research staff of an organisation. Overall, for the field of observation adopted in this study, this meant 56% more subjects in the CWTS dataset than in the ORP dataset, which draws on guaranteed data from the MUR. One of the reasons for this would be that the CvE approach, which underlies the CWTS dataset, attributes all researchers to an organisation when these have prevalently indicated the relative affiliation in signing their scientific publications, independently of their effective position within the organization. In this way, doctoral and post-doctoral students, post-doctoral fellows, collaborators and a range of other individuals who would not be eligible for evaluation in an official national evaluation exercise also end up on a university's list of researchers.

It should also be considered that the CvE algorithm, like any unsupervised approach, tends to favour precision over recall; in particular, the publication oeuvre of an author can be split over multiple "clusters" if not enough proof is found for joining publications together. This means that the actual value of the over-representation of an organisation's research staff in the CWTS dataset is lower than the figure measurable by direct comparison with the ORP data. At an overall level, therefore, that +56% represents an upper bound of the actual incidence of non-faculty personnel in the CWTS dataset.

Having said this, the scores and ranks recorded in the two compared modes show a significant and rather high correlation: at the overall level, Pearson and Spearman coefficients, respectively, are 0.813 and 0.686. At the area level, the values are never below 0.5 with peaks in Chemistry (0.851; 0.824). The only critical area is Psychology, however this is an area present only in a small number of assessed universities, at nine. Still, the overall correlation covers significant jumps at local level: comparing the CWTS assessment to the benchmark, in the ranking of all 69 assessed universities, nine show deviations of two quartiles, better or worse.

Among the drivers of these deviations, certainly the greatest weight goes to the different number of observations between the two datasets. Empirically, it emerges that percentage deviations in FSS for universities are significantly and negatively correlated with percentage deviations in numerosity between the two datasets.

If we assume that the over-representation of the CWTS dataset with respect to ORP depends largely on the CWTS inclusion of "non-faculty" personnel, we can deduce that on average, these personnel perform less well. More importantly, it can also be concluded that an evaluation conducted by means of the CvE approach, although there could be exceptions, would generally penalise universities (and areas) with a higher proportion of non-faculty research-active personnel.

Obviously this effect, which is certainly significant, must be discounted against the intrinsic limits of CvE. Notably, as mentioned above, such an algorithm can in some cases attribute a researcher's scientific production to two (or more) distinct "clusters", especially in the presence of a scientific production characterised by heterogeneous and highly differentiated bibliometric metadata. The impact of such "splitting" on the outcomes of the comparative evaluation of universities should, however, be very limited, since the splitting cases should be evenly distributed and not focus on researchers from one organisation over those from others. And, it is worth remembering, the literature in any case indicates CvE as the best performing of such unsupervised algorithms (Tekles & Bornmann, 2020).

While waiting for policy makers to take action towards national and international systems for collecting input data, which would enable bibliometricians to carry out what the same policy-makers are ever more insistently demanding, practitioners could consider using the CWTS data as in this current paper. In particular, the methodology described here should make it possible for others to replicate the comparative analyses in the frameworks of their interest (national or international), simply by processing the output of the CvE algorithm in an appropriate manner, in particular by considering the relative institutions' official URL domains. A notable side benefit of this would be that with this, practitioners now have a precise measure of the extent of distortions inherent in any evaluation exercises using unsupervised algorithms.

And, for policy makers, knowing the extent of performance measure distortions could be useful in deciding whether to invest in development of databases of national research personnel, or to settle for the less precise assessments.